# Decision under ambiguity: Effects of sign and magnitude.

Keigo Inukai[1], Taiki Takahashi[2]

[1] Laboratory of Social Psychology, Department of Behavioral Science, Faculty of Letters, Hokkaido University
N.10, W.7, Kita-ku, Sapporo, 060-0810, Japan
TEL +81-11-706-3057, FAX +81-11-706-3066

[2] Laboratory of Social Psychology, Department of Behavioral Science, Faculty of Letters, Hokkaido University
N.10, W.7, Kita-ku, Sapporo, 060-0810, Japan
TEL +81-11-706-3057, FAX +81-11-706-3066
taikitakahashi@gmail.com(corresponding author)

Abstract

Decision under ambiguity (uncertainty with unknown probabilities) has been attracting attention in behavioral and neuroeconomics. However, recent neuroimaging studies have mainly focused on gain domains while little attention has been paid to the magnitudes of outcomes. In this study, we examined the effects of the sign (i.e. gain and loss) and magnitude of outcomes on ambiguity aversion and the additivity of subjective probabilities in Ellsberg's urn problem. We observed that (i) ambiguity aversion was observed in both signs, and (ii) subadditivity of subjective probability was not observed in negative outcomes.

1. Introduction

Decision under uncertainty with known probabilities has been extensively studied in both behavioral and neuro- economics. For instance, recent neuroimaging studies have shown that one's parameter in Prelec's probability weighting function is associated with neural activation in the anterior cingulated cortex (Paulus et al., 2006), and Kahneman-Tversky's framing effects in decision under probabilistic uncertainty across gain-loss domains have been found to be negatively related to the orbitofrontal activities (De Martino et al., 2006). We have also examined the neurochemical modulation of probabilistic choice (Ohmura et al., 2005), and the role of information-theoretic uncertainty in probabilistic choice (Takahashi et al., 2007). Moreover, several studies have examined decision under uncertainty with unknown probability (ambiguity) (Camerer & Weber, 1992). For instance, McCabe's group performed a PET (positron emission tomography) study across gain-loss domains (Smith et al., 2002), Camerer's group has demonstrated that ventromedial-amygdala circuits are neural circuitry of general uncertainty evaluation (Hsu et al., 2005), and Huetell's group has reported that risk and ambiguity preferences may be represented in the posterior parietal cortex and lateral prefrontal cortex, respectively (Huettel et al., 2006).

However, to date, no study has systematically examined the effects of sign and magnitude of outcomes in a large sample population. In this study, we examined the roles of sign and magnitude in decision under ambiguity in terms of ambiguity aversion and the additivity of subjective probability of ambiguous outcomes, by utilizing Ellsberg's urn problem with different magnitudes and signs of outcomes (large, medium, and small gain and loss) in university students. Because the effects of sign and magnitude have been extensively studied in intertemporal and probabilistic choice (Estle et al., 2006; Frederick et al., 2002), our present study may also help elucidate distinct psychological processes for decision-making under ambiguity and risk, and over time.

2. Methods

A total of 122 university students participated in the present study. In order to examine subjects' attitudes towards ambiguity and risk, the Ellsberg's urn problem with hypothetical outcomes was employed. It should be noted that employing real monetary payoffs is difficult for studies on decision-making in loss domains. Subjects were instructed in the following manner (the example of the small gain condition is presented in detail). It is to be noted that ¥1000 (JPY) is approximately equivalent to US $10.

[Imagine that an urn contains 90 balls: 30 are red; of the rest, some are blue and some are green. One ball is to be drawn at random form the urn.

First question:
You are asked to choose between the following options:
Option A. Receive ¥1,000 if the ball is red.
Option B. Receive ¥1,000 if the ball is blue.

Second question:
You are asked to choose between the following options:
Option C. Receive ¥1,000 if the ball is red or green.
Option D. Receive ¥1,000 if the ball is blue or green.]

In the negative outcome tasks, we replaced "receive" with "lose" in order to examine subjects' attitudes towards ambiguity in loss domains. To examine the effect of magnitudes of outcomes, three different sizes (i.e., ¥1,000, ¥10,000, ¥100,000) of gains and losses were presented (i.e. a total of six conditions were examined). It is important to note that (i) if a subject prefers option A (risk) over option B (ambiguity), s/he is ambiguity aversive, and (ii) if a subject prefers option A over option B and, simultaneously, prefers option D over option C, subjective probabilities estimated by the subject for possible ambiguous outcomes are subadditive (Camerer & Weber, 1992). Therefore, according to these two criteria, we examined subjects' ambiguity aversion and additivity of subjective probability in small, medium, and large gains and losses. In order to examine whether subjective probabilities are subadditive or not, we utilized McNemar tests for within-individual comparisons of choices between risk (option A and option D) and ambiguity (option B and option C). Significance levels were set at 0.05 throughout.

3. Results

[Figure 1 inserted here]

Ambiguity aversion
We first analyzed subjects' choices for the first question, in order to examine whether or not subjects were ambiguity aversive in each sign and magnitude. Across all signs and magnitudes, most subjects chose risky options (option A, options under uncertainty with known probabilities), rather than ambiguous options (option B, options under uncertainty with unknown probabilities). This indicates that subjects were ambiguity aversive in loss domains as well as gain domains. The proportion of subjects choosing risky options did not significantly differ across signs and magnitudes of outcomes.

[Table 1 and 2 inserted here]

Subadditivity of subjective probability

Next, we analyzed the additivity of subjective probability of ambiguous outcomes. It should be noted that if subjective probabilities of ambiguous outcomes are subadditive, subjects will choose option A and option D in the gain domains, but will choose option B and option C in loss domains. The proportion of subjects' choices in gain and loss are presented in Tables 1 and 2, respectively. As we can see, subjective probabilities for ambiguous gains were subadditive ($p<0.001$, for all gains); while subjective probabilities for ambiguous losses were not subadditive, indicating that subjects might estimate probability of ambiguous loss in a non-subadditive (additive or superadditive) manner. Moreover, for medium loss, subjects' choices were inconsistent in terms of ambiguity aversion, in that they more markedly chose the ambiguous option in the second question, while significantly avoiding ambiguity to a great extent in the first question (Table 2B).

4. Discussion

This study systematically examined the effects of magnitudes and signs of outcomes on decisions under ambiguity. We observed that (i) subjects are ambiguity aversive in both gains and losses across all magnitudes, and (ii) subjective probabilities of ambiguous outcomes are subadditive in gains but not subadditive in losses.

4.1 Magnitude effects in decision under uncertainty

With a few exceptions (Estle et al., 2006; Ohmura et al., 2005; Smith et al., 2002), previous studies on intertemporal and probabilistic choices, mostly focusing on gain domains have reported that people discount delayed large gains less steeply than small gains; while large unlikely gains are more strongly devaluated than small unlikely gains (Frederick et al., 2002; Estle et al., 2006). Our current findings indicate that there is no significant magnitude effect in decision-making under ambiguity in terms of choice between ambiguous and probabilistic outcomes (relative subjective values between outcomes with risk and ambiguity). However, when certain outcomes which are subjectively equivalent to ambiguous outcomes are examined, it is possible for magnitude effects appear. Furthermore, magnitude effects in loss domains have not been extensively studied even in intertemporal and probabilistic choices; these points should be examined in future studies.

4.2 Sign effects in decision under uncertainty

We observed that there was no sign effect on ambiguity aversion in the first question of the Ellsberg's urn problem, consistent with previous reports (Inukai & Takahashi., 2006; Smith et al., 2002) and in contrast to intertemporal and probabilistic choices. In intertemporal choice, it is known that gains are more steeply discounted than losses; while in decision under risk, it is well-established that people are risk-aversive in gain frames, but risk-seeking in loss frames, which has been associated with the amygdala activation by utilizing functional magnetic resonance (fMRI) imaging in a recent neuroeconomic study (De Martino et al., 2006). Therefore, future neuroeconomic studies may be able to identify distinct neural processes between decision under ambiguity, and intertemporal and probabilistic choices.

4.3 Additivity of subjective probability in decision under uncertainty

Additivity of probability in decision under risk (probabilistic uncertainty) and patience in intertemporal choice (i.e. a discount factor) has been extensively studied (Read et al, 2003). In probabilistic choice, it is well-established that people overestimate small probabilities while underestimating large probabilities. The behavioral economist Prelec has axiomatically derived the probability weighting function, with which it can be stated that large probabilities are subadditive; while small probabilities are superadditive. A recent neuroeconomic study demonstrated that distortion of probability weighting is negatively associated with the activation of the anterior cingulate cortex; while no neuroimaging study has demonstrated the neural correlate of subadditive time-discounting (Paulus et al., 2006). Regarding the neural processing underlying ambiguity aversion, Camerer's group has shown that in gain domains ambiguity is associated with stronger activation of cortico-limbic circuitry for general uncertainty rather than risk. Future studies utilizing PET/fMRI should examine whether or not different degrees of neural activation in the cortico-limbic neural circuits are apparent. This point may help to understand the mediating effect of neural processes in "choice bracketing" (Read et al., 1999) in decisions under ambiguity. In the current study, it was observed that, for medium loss, subjects' choices were inconsistent in terms of ambiguity aversion. Future neuroimaging studies are needed to elucidate the neural processing behind inconsistent attitudes toward ambiguity in loss domains.

Table 1A Ellsberg's urn problem with small gain

| +¥1,000 | | Second question | |
|---|---|---|---|
| | | C:Ambiguity (red and green) | D: Risk (blue and green) |
| First question | A: Risk (red) | 16.7% | 53.3*% |
| | B: Ambiguity (blue) | 11.7% | 18.3% |

Each cell indicates the proportion of choices made by subjects. McNemar test revealed that subjective probability for ambiguous small gain was subadditive, i.e., subjects significantly prefer risky over ambiguous gains in both questions ($\chi_0^2$=16.7, p<0.001).

Table 1B Ellsberg's urn problem with medium gain

| +¥10,000 | | Second question | |
|---|---|---|---|
| | | C:Ambiguity (red and green) | D:Risk (blue and green) |
| First question | A: Risk (red) | 18.3% | 53.3*% |
| | B:Ambiguity (blue) | 13.3% | 15.0% |

Each cell indicates the proportion of choices made by subjects.

McNemar test revealed that subjective probability for ambiguous small gain was subadditive, i.e., subjects significantly prefer risky over ambiguous gains in both questions ($\chi_0^2$=14.4, p<0.001).

Table 1C Ellsberg's urn problem with large gain

| +¥100,000 | | Second question | |
| --- | --- | --- | --- |
| | | C: Ambiguity (red and green) | D: Risk (blue and green) |
| First question | A: Risk (red) | 21.3% | 54.1*% |
| | B: Ambiguity (blue) | 9.8% | 13.1% |

Each cell indicates the proportion of choices made by subjects.

McNemar test revealed that subjective probability for ambiguous small gain was subadditive, i.e., subjects significantly prefer risky over ambiguous gains in both questions ($\chi_0^2$=18.7, p<0.001).

Table 2A Ellsberg's urn problem with small loss

| – ¥1,000 | | Second question | |
|---|---|---|---|
| | | C: Ambiguity (red and green) | D: Risk (blue and green) |
| First question | A: Risk (red) | 35.0% | 30.0% |
| | B: Ambiguity (blue) | 21.7% | 13.3% |

Each cell indicates the proportion of choices made by subjects.

McNemar test revealed that subjective probability for ambiguous small loss was not subadditive, i.e., subjects did not consistently avoid risky losses, in comparison to ambiguous losses, in both questions ($\chi_0^2$=0.8, p=0.37).

Table 2B Ellsberg's urn problem with medium loss

| – ¥10,000 | | Second question | |
|---|---|---|---|
| | | C: Ambiguity (red and green) | D: Risk (blue and green) |
| First question | A: Risk (red) | 43.3*% | 33.3% |
| | B: Ambiguity (blue) | 11.7% | 11.7% |

Each cell indicates the proportion of choices made by subjects.

McNemar test revealed that subjects significantly avoided ambiguous than risky losses in the first one-color question but avoided risky losses more strongly in the second two-color question ($\chi_0^2$=6.26, p=0.01<0.05), indicating that their attitude towards ambiguity was significantly inconsistent. Anyway, it can be said that subjective probability for ambiguous medium loss was not subadditive, i.e., proportion of cell (2,1) = 11.1 was not significantly larger than other proportions.

Table 2C Ellsberg's urn problem with large loss

| – ¥100,000 | | Second question | |
|---|---|---|---|
| | | C:Ambiguity (red and green) | D:Risk (blue and green) |
| First question | A:Risk (red) | 43.3% | 28.3% |
| | B:Ambiguity (blue) | 13.3% | 15.0% |

Each cell indicates the proportion of choices made by subjects.

McNemar test revealed that subjective probability for ambiguous large loss was not subadditive, i.e., subjects did not consistently avoid risky losses, in comparison to ambiguous losses, in both questions ($\chi_0^2$=3.24, p=0.072).

Figure 1: Ambiguity aversion in the Ellsberg's urn problem in the first question in both signs (gain and loss).

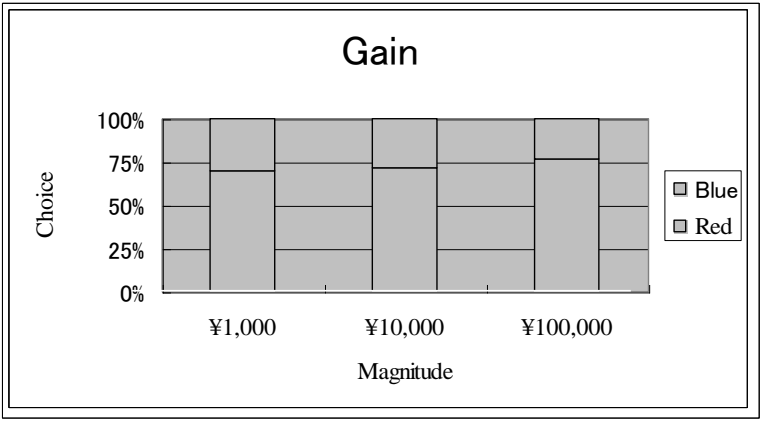 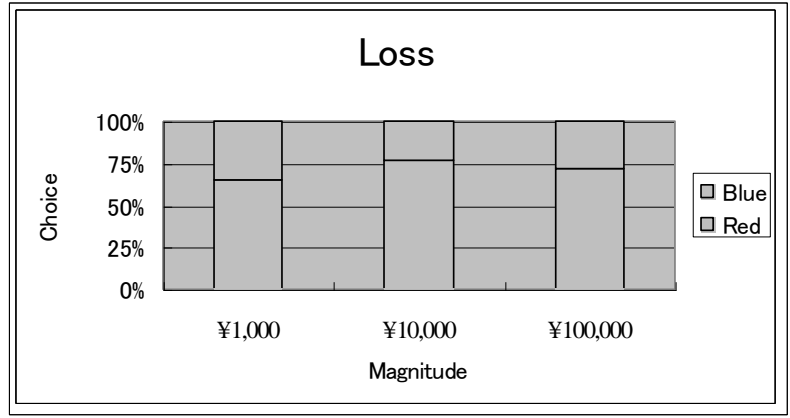

Subjects significantly avoided ambiguous options in both gains and losses.